\documentclass[12pt]{article}

\usepackage[utf8]{inputenc}
\usepackage[T1]{fontenc}

\usepackage[margin=1in]{geometry}

\usepackage{xcolor}
\usepackage{parskip}         
\usepackage[hidelinks]{hyperref}
\usepackage{xurl}            
\usepackage{enumitem}        
\usepackage{setspace}

\title{\vspace{-1.5em}Democracy Needs Reach: Political Equality, Online Speech, and Algorithmic Recommendation}
\author{Étienne Brown\\[2pt] \small University of Ottawa}
\date{}

\newenvironment{hangingbib}
  {\small\begin{list}{}{%
     \setlength{\leftmargin}{1.5em}
     \setlength{\itemindent}{-1.5em}
     \setlength{\itemsep}{0.6em}
     \setlength{\parsep}{0pt}
  }}
  {\end{list}}

\begin{document}

\maketitle
\vspace{-2em}
\begin{center}
\textit{Preprint. Please cite the published version in \textit{Ethical Theory and Moral Practice}.}
\end{center}
\vspace{1em}

\begin{abstract}
Within democracies, the capacity to influence political outcomes through speech depends not only on the right to express oneself, but also on the opportunity to reach relevant audiences. In this paper, I argue that the unequal distribution of algorithmic reach on social media platforms undermines equality of opportunity for political influence (EOPI), which is a central democratic ideal. Drawing on Niko Kolodny's work, I contend that current recommendation algorithms create and perpetuate informal inequalities by concentrating attention among a small minority of already-amplified speakers while systematically marginalizing others. To address this problem, I propose recommendation floors as a mechanism for equalizing political speech. To help users achieve meaningful participation, each verified account would receive guaranteed minimum recommendation for up to a limited number of political posts per week. Although this measure represents one component of the structural reforms needed to move the digital public sphere closer to democratic ideals, it offers a feasible pathway to reducing informal inequalities in political influence online.
\end{abstract}

Let me start by painting a familiar scene. In the middle of a crowded
pedestrian street, a woman stands on a soapbox.\footnote{I would like to thank Luise Müller, Henrik Kugelberg, and Zoe
  Phillips Williams for their detailed comments on previous versions
  of this article. For helpful feedback, I am also grateful to Jeffrey
  Howard, Hanna Kiri Gunn, Seth Lazar, Erin Miller, Chloé Bakalar,
  Regina Rini, Jonathan Quong, Mark Schroeder, J. L. A. Donohue, Aksel
  Sterri, Joseph Heath, Samuel Dishaw, Gabriele Badano, Susan Brison,
  Barrett Emerick, Mihaela Poppa-Wyatt, Marianna Ganapini, Maria
  Brincker, Kate Norlock, Lucy McDonald, Rob Reich, Elizabeth
  Edenberg, and Mark Satta.} In one hand, she
holds a cardboard sign with verse numbers; in the other, a megaphone
through which she loudly proclaims that ``Jesus saves.'' Around her,
passers-by quicken their pace. Some have a meeting to attend; others are
not especially worried about the fate of their soul. Despite her best
efforts, the preacher primarily speaks to the pavement. Of course, we
all know she would prefer an audience. She needs one to save souls. Like
most, she believes in a cause and longs for a conversation. A nod, a
rebuttal, or even an insult would be welcomed. It would remind her that
her utterances reach the minds of others.

Still, the thought that the preacher is not expressing herself freely
would not cross any observer's mind. She has a soapbox, a cardboard
sign, and a megaphone. No one is preventing her from speaking. She is
exercising her democratic right to free speech. In a sense, her tale is
a sad one; in another, it is a story of success. The preacher wanted to
preach in \emph{this} crowded pedestrian street at \emph{this} chosen time with
\emph{these} instruments, and she did. When passers-by ignore her, she has
already won the first battle. She might not have been listened to, but
she has been heard, and this is not an accomplishment one should take
for granted.

To see this, imagine that our preacher no longer speaks in the analogue
public sphere. Instead, she posts her sermons on a mainstream social
media platform. Metaphorically speaking, whether the street on which she
preaches is crowded is no longer up to her. In the digital public
sphere, she might also be unable to keep her soapbox, cardboard sign,
and megaphone. From now on, the reach of her speech will be determined
by recommendation algorithms that she does not control. If a recommender
predicts that her posts will not be engaging, they will not be widely
distributed. Alternatively, her speech might be submerged in a sea of
content shared by users with greater means to augment their own.
Perhaps, for instance, these users have access to a prebuilt audience of
loyal followers or to funds that allow them to purchase a much larger
share of the digital public's collective attention. (Miller 2021)

If the voices of others systematically bury the preacher's speech, do we
have reasons to worry? In a democratic society, is it a problem if the
reach of some political speech is very low? In this paper, I suggest
that it is. Specifically, I argue that lowly amplified online political
speech poses a problem for democracy, as it erodes equality of
opportunity for political influence (EOPI). From the perspective of
democratic equality, there are also compelling democratic reasons to
pressure social media platforms into implementing measures that would
equalize the amplification of such speech. My discussion proceeds as
follows. In section 1, I anchor my argument in considerations about the
relationship between speech, reach, and uptake. In section 2, I present
the core argument of my paper. Drawing from Niko Kolodny's egalitarian
critique of social hierarchy, I contend that opportunities for political
influence are unfairly distributed amongst political speakers in the
digital public sphere. I then propose a practical measure that would
help equalize the amplification of political speech at the platform
level: recommendation floors that would guarantee a minimum level of
``impressions'' per user for qualifying political content (section 3).
Lastly, I respond to three objections according to which (i) private
social media companies have a stringent proprietary right to exclude
political speakers at will, (ii) equalizing amplification runs contrary
to users' ability to maintain agency over what content they are exposed
to, and (iii) recommendation floors would unduly constrain the economic
activity of social media platforms and their users (section 4).

Before I begin, a few remarks are in order. First, my primary aim in
this article is to contribute to a broader philosophical reflection on
the set of ethical and political values with which recommendation
algorithms should be aligned. (Stray et al.~2022) In this context, I
suggest that democratic equality should be included in this set. Such a
philosophical reflection seems important to me at a time when the
European Union is both actively regulating social media platforms and
funding large-scale research projects on alternative recommendation
algorithms. Second, my reflection aligns with a growing body of research
that highlights the moral and political problems arising from attention
maldistribution.\footnote{For instance, scholars have recently argued that some forms of
  celebrity involvement in politics threaten the legitimacy of
  existing democracies (Archer et al.~2020). Smith and Archer (2020)
  also suggest that giving a person less attention than they are due
  is a form of epistemic injustice.} Here, a common concern is that some members of the
\emph{demos} are excluded from online political discussions or, at the very
least, not given their fair share of collective attention. This includes
those who do not have money to spend on means to amplify their speech,
those who do not have countless hours to spend on social media, those
who do not possess the required \emph{know-how} to have their voices heard in
digital town squares, and those whose voices are not heard as loudly as
those of others because of stereotypes and prejudices. Third, my
reflection deliberately moves from ideal theory to non-ideal theory.
After a discussion of EOPI, I analyze current platforms' practices and
show how far they fall short of egalitarian ideals. I then propose a
practical step toward EOPI.\footnote{In doing so, I follow other applied philosophers who have begun
  reflecting upon the implications of EOPI for the digital public
  sphere. According to Bousquet (2025), for instance, EOPI entails
  that social media platforms should not be allowed to engage in
  viewpoint-based content moderation.} As I explain, equalizing the reach of
online political speech is feasible within our current digital media
ecosystem, as it would not require platforms to develop new technical
capacities. Of course, social media platforms are not likely to
implement algorithmic reforms simply because they are concerned about
democracy. When I claim that platforms \emph{should} implement recommendation
floors\emph{,} I mean the same as when other scholars claim, for instance,
that platforms \emph{should} combat medical misinformation regardless of
their current incentives to do so. Of course, it may well be that
regulation is needed to align platform practices with democratic ideals.
Before we regulate, however, we should have a clear representation of
the digital public sphere we want rules to help us create. My objective
in what follows is to contribute to this reflection.

\section*{1. Speech, Uptake, and the Value of Reach}\label{speech-uptake-and-the-value-of-reach}

Let us begin with the idea that the pursuit of speech-related interests
requires more than just speech alone. Effectively disclosing ourselves
and engaging in meaningful relationships with others requires \emph{uptake},
that is, the understanding of the meaning and of the force of locutions.
Unless our interlocutors can reliably discern the meaning of our
utterances, our lives are likely to go very poorly. Matteo Bonotti and
Jonathan Seglow (2022) nicely capture this thought in their recent
relational defense of freedom of expression. As they explain:

\begin{quote}
\ldots the negative right to free speech and the opportunity to exercise
it are analytically distinct. But we need to consider the meaning of a
free speech right from the perspective of a potential speaker who
wishes to communicate something to others, or to influence them, or
make a request of them, or to engage in deliberation with them, and so
on {[}\ldots{]} The point of free speech is to address others and possess
some traction in doing so; a person's speech, to have value for them,
needs to have the potential for uptake. Without this, free speech, for
some at least, is a liberty without worth. (517)
\end{quote}

This passage helps us see that speakers must be \emph{heard} if they are to
pursue their speech-related interests successfully. If your right to
free expression merely allows you to shout in an empty room, it will be
of very little value to you. This is because speech is rarely an end in
itself but typically an instrument to an end. Note also that such
considerations do not implicate political speech in particular. To
pursue fundamental interests, such as being known and respected for the
individual one is, speakers must have the opportunity to engage in
successful communication on a wide range of topics. (Shiffrin 2014,
88-89) As Shiffrin explains, defenses of free speech that build on a
general conceptualization of the communicative interests of persons make
``no important distinction, at the foundations, between speech about
aesthetics, one's medical condition and treatment, one's regard for
another, one's sensory perceptions, the sense or lack thereof of the
existence of a God, or one's political beliefs.'' (92-93) All these
communications provide listeners with the information necessary to
appreciate the speaker ``as a distinct individual and to forge a fuller
human relation with her.'' (93)

If all meaningful speech requires access to prospective audiences, why
restrict the equalizing measures I describe in section 3 to \emph{political}
speech? As we will see, a central reason is that democracy requires
equal opportunity for political influence, not equal opportunity for
influence \emph{tout court}. Erin Miller's recent study of amplified speech
also helps answer this question. According to her perspective, what
matters for speakers' interests is that they receive an audience of
sufficient size. ``To engender respect, Miller writes (2021, 52), our
self-expression needs a certain type of audience---but not necessarily a
large one.'' This is because ``a small community of supporters is for most
people sufficient to establish self-respect.'' (52) Beyond a certain
point, greater amplification provides diminishing marginal returns to
speakers. Applying Miller's reasoning to the digital public sphere, we
can say that speakers need not have all their social media posts seen by
hundreds of users to pursue fundamental interests such as self-knowledge
and self-respect. Having the opportunity to form a small online
community on a platform like Reddit or Facebook is likely to suffice.
However, there is a speech-related interest to which Miller's
diminishing marginal return argument does not apply: our democratic
interest in influencing the political process and contributing to the
formation of public opinion through expression. In a democratic society,
speakers compete for the scarce attention of fellow citizens, and the
amount of influence they wield partly depends on the share of collective
attention captured by their political adversaries. If one's political
adversaries command a much larger share of the democratic public's
attention, there is an important sense in which I am not in an equal
relationship with them. While such considerations help explain why reach
is valuable to political speakers, they also raise daunting questions.
What kind of equality do democratic citizens owe one another? Should
they endeavor to create a society in which lack of reach does not
prevent anyone from meaningfully participating in political discussions?
If so, how can they do so? In the next sections, I turn to these
questions.

\section*{2. Democracy Needs Reach}\label{democracy-needs-reach}

Recently, Niko Kolodny (2014) has argued that, in a democratic context,
citizens have a fundamental interest in being provided with an equal
opportunity to influence the political decision-making process.
According to his perspective, disparities of influence fall into two
broad categories: inequalities of formal influence and inequalities of
informal influence. As their name suggests, inequalities of formal
influence relate to formal procedures. Here, voting is the most obvious
example. For democratic citizens to have equal influence over the
political decision-making process, universal franchise must be
implemented, and votes must be equally weighted. By way of contrast,
informal opportunity ``consists roughly in the availability of resources,
such as wealth and leisure, to apply to the legal or procedural
structure to acquire information or influence the votes of others.''
(289-290) If you and I have equal votes, but I am so powerful that I can
manipulate voters into voting as I want them to, then we are political
equals only in an empty sense. Although inequalities of formal
influence, such as unequal voting rights, represent an especially
salient case of injustice, Kolodny argues that social equality requires
the elimination of both formal and informal inequalities of influence.
As he explains (2014, 332):

\begin{quote}
As far as social equality is concerned, inequalities of influence over
political decisions resulting from informal conditions are (at least
in the first instance) no less threatening than inequalities of
influence over political decisions built into formal procedures.
\end{quote}

If A has unlimited access to campaign spending or lobbying and can, as a
result, systematically ensure that political outcomes favor his
interests rather than those of B, then B stands in a relationship of
inferiority to A. If so, knowing that A and B have equal voting rights
will be of little consolation to B. Kolodny sums up this idea by writing
that ``As far as Equal Influence {[}\ldots{]} is concerned, influence is
influence, whether formal or informal.''

Consequently, EOPI not only requires that formal procedures be put in
place (e.g., equal voting rights) but also the fair distribution of
resources that enable people to influence political outcomes. In this
section, my central suggestion is that \emph{reach}, that is, the capacity to
be heard and affect the minds of others through speech, is one of those
resources. Ronald Dworkin (2011, vii-viii) forcefully underlines this
point when he writes:

\begin{quote}
A majority decision is not fair unless everyone has had a fair
opportunity to express his or her attitudes or opinions or fears or
tastes or presuppositions or prejudices or ideals, not just in the
hope of influencing others, though that hope is crucially important,
but also just to confirm his or her standing as a responsible agent
rather than a passive victim of collective action {[}\ldots{]} It may be
objected in most democracies that right now, the right to free speech
has little value for many citizens, ordinary people with no access to
great newspapers or television broadcasts have little chance to be
heard. This is a genuine problem. It may be that genuine free speech
requires more than just freedom from legal censorship {[}\ldots{]} we must
try to find other ways of providing those without money or influence a
real chance to make their voices heard.
\end{quote}

Here, Dworkin argues that although society must give those without money
or influence a real chance to be heard, censorship is never a
justifiable means to equalize influence. To ensure that a subset of
speakers has an equal opportunity to influence others\textquotesingle{} minds, we should
not censor racists and misogynists, even if their speech has silencing
effects. Yet this thought is perfectly compatible with the idea that
voice should be redistributed in other ways. Perhaps, for instance,
grassroots political candidates who face wealthy opponents should
receive public funding if we want them to stand a chance in the
electoral process. Or perhaps, as I will suggest in the next section,
recommender systems should be designed to reduce disparities in
political influence among speakers in the digital public sphere. The
point here is that there are properly democratic reasons to avoid
creating a public sphere in which elites have great means to amplify
their speech while others are barely heard.

Furthermore, even those who do not and perhaps will never make use of
this opportunity should be given it if they are to relate to others as
equals. For instance, we would not find it acceptable to implement a
plural voting scheme that gives less weight to the votes of people who
rarely go to the polls. If we did, not only would we deprive them of
their opportunity to influence, but we would also show them unequal
consideration. Indeed, we would be treating them as unworthy of the
powers we believe we deserve. This requirement of equal standing easily
applies to political speech. Since voice is a central instrument of
influence, those who are actively trying to shape the democratic
decision-making process have a political interest in being provided with
an equal opportunity to be heard. However, those who do not actively
engage in political speech \emph{also} have an egalitarian interest in being
provided with that opportunity, as it confirms their standing as equal
members of the \emph{demos}.

Interestingly, democratic institutions frequently recognize the
importance of reach for political speech. By way of example, American
``constitutional doctrine implies that speech rights are meaningless
without the opportunity to reach an audience.'' (Miller 2021, 10) In
Kovacs v. Cooper, Supreme Court Justices also underlined that ``The right
of free speech is guaranteed every citizen that he may reach the minds
of willing listeners and to do so there must be opportunity to win their
attention.''

Are there reasons to believe that the digital public sphere is plagued
by inequalities of reach? I believe there are. On social media
platforms, speech typically travels through a mix of network propagation
and algorithmic recommendation. (Narayanan 2023) When information
propagates through a network, it cascades ``as long as users who see it
choose to further propagate it.'' For example, network propagation is at
play when X or Bluesky users see a piece of content because another user
they follow has posted (or reposted) it. What made mainstream social
media platforms so successful, however, is not network propagation but
their choice to mix it with algorithmic recommendation. On such
platforms, user-generated content is distributed by a recommender
system, that is, an algorithm ``that takes a large set of items and
determines which of those to display to users.'' (Thorburn, Bengani, and
Stray 2022) Currently, the central function of these algorithms is to
optimize for engagement, i.e., to rank and display items predicted to be
engaging for users. Conversely, content that is estimated not to be
engaging receives a low score and is not widely distributed.

That all democratic speakers currently have unequal opportunities to
have their voices heard in the public sphere seems quite plausible.
First, and most importantly, empirical analyses of current biases in
recommendation algorithms reveal that attention is highly concentrated
on social media platforms. One phenomenon that can be observed is the
``rich-get-richer, poor-get-poorer'' dynamic. As Narayanan (2023)
explains, ``Those who already have a high reach, whether earned or not,
are rewarded with more reach.'' Using Gini coefficients to measure
attention inequality on Twitter, for instance, Zhu and Lerman found that
``the top 20\% of Twitter users own more than 96\% of all followers, 93\% of
retweets, and 93\% of mentions.'' This means that ``the vast majority of
users do not receive any attention. Even more dramatically,''the top 1\%
of users get far more attention than the bottom 99\% combined!'' Moreover,
those who are already poor in attention are likely to lose followers
over time. In this case, exorbitant inequalities of reach systematically
marginalize speakers by effectively drowning them out of sight or
hearing.

Second, platform owners can advantage their own speech. In 2023, for
instance, Elon Musk ordered X's software engineers to modify the
platform's recommendation algorithms to ensure that his posts receive
``previously unheard-of promotion of his tweets to the entire user base.''
(Schiffer and Newton 2023) Third, wealthy speakers can purchase
amplification through advertisements or paid features such as Facebook's
``boost'' or X's ``promote.'' Monthly subscription services like X Premium
(previously Twitter Blue) also provide users with ``prioritized rankings
in conversations and search,'' for instance, by ensuring that their
replies rank ``closer to the top'' (Twitter, 2023). Fourth, inequalities
in leisure time and digital literacy affect reach. Although social media
platforms have undoubtedly democratized access to the public sphere,
some speakers still lack the resources to make full use of this
opportunity. Acquiring a large number of followers is a lengthy process
that arguably requires a form of knowledge that some users are more
likely to acquire, given their educational history, age, and
socioeconomic status. Indeed, social scientists are increasingly
concerned that age and poverty are driving factors of exclusion in the
digital public sphere. According to an egalitarian framework, however,
undeserved inequalities resulting from money, time, digital literacy,
and a lack of control over algorithms are incompatible with democratic
equality when they translate into inequalities of political influence.

These considerations raise difficult questions. If unfair inequalities
of influence plague the digital public sphere, how can we ensure they
don't? Theoretically, what conditions must be met to achieve equality of
opportunity for political influence? Practically, how can we do so? I
believe we can answer the theoretical question by applying Rawls's
concept of fair equality of opportunity to social media platforms.
According to this perspective, two people with the same ambition to
communicate their political views to an audience and the same level of
talent should have the same likelihood of capturing that audience's
attention. Here, ``talent'' is understood in a specific sense; whether a
speaker has it depends on whether prospective audiences want to allocate
attention to their speech. At the platform level, however, this
philosophical interpretation of equality of opportunity for political
influence raises practical challenges. It is difficult -- and perhaps
impossible -- to determine whether two political speakers with the same
ambition to share their views have unequal reach due to unfair
inequalities (e.g., lack of leisure time). Digital platforms cannot
reasonably be expected to know this, and a world in which they did would
likely raise significant privacy concerns. Moreover, determining whether
prospective audiences would want to allocate attention to a user's
speech if they were exposed to it in the first place is also
challenging. For practical reasons, one might therefore be tempted to
abandon the project of equalizing political influence in the digital
public sphere. Yet my suggestion is that the cost of this practical
pessimism -- a state of affairs in which our democratic ideals are
celebrated without being met -- is too high.

To overcome this challenge, I propose applying a \emph{charity principle} to
online political speakers. In a context where we have serious reasons to
believe that unfair informal inequalities affect the reach of online
political speech, we should refrain from assuming that lowly amplified
political speakers deserve it and that others are uninterested in their
views. Consequently, we should provide them with a greater opportunity
to be heard. This principle certainly carries the risk of amplifying
political speakers whose low political influence is not caused by unfair
inequalities. From the perspective of democratic equality, however,
taking the opposite risk (as we currently do) is worse. Of course, how
we can provide online political speakers with a greater opportunity to
be heard is not obvious. Let us then see how.

\section*{3. Equalizing the Algorithmic Recommendation of Political Speech}\label{equalizing-the-algorithmic-recommendation-of-political-speech}

If the preceding remarks are plausible, egalitarians have democratic
reasons to seek to equalize the amplification of political speech on
social media platforms. Before considering how this could be done, let
me first address a question about platform \emph{responsibility}. If
opportunities to influence the democratic process are unequally
distributed in the digital public sphere, can democratic citizens demand
that non-governmental actors, such as social media platforms, temper
inequalities of reach? Even if we worry about such inequalities, the
conclusion that social media platforms have a moral duty to remedy this
problem does not immediately follow. Other social actors may have a
greater capacity to address this problem or bear primary responsibility
for its creation. In a recent discussion of communicative justice,
however, Seth Lazar (2025) argues that users and non-users alike can
legitimately demand that platforms attempt to ameliorate pathologies
plaguing the digital public sphere.~In his view, the very fact that
immense platforms have intentionally designed and currently govern new
communication spaces makes them responsible for the pathologies that
affect them, including misinformation, harassment, polarization, and
silencing.

In my view, Lazar's reasoning also applies to the problem of unequal
opportunity to reach. Social media companies wield immense power in
determining whose voices are amplified and whose are reduced in the
digital public sphere. While users' reactions drive content
distribution, how recommender systems weigh these reactions is entirely
determined by platforms. For example, social media companies regularly
decide what ``Likes,'' ``Reshares,'' and ``Comments'' are worth. At \emph{t\textsuperscript{1}}, a
recommender might be programmed to weigh one ``Comment'' as ten ``Likes.''
If this yields undesirable results -- think of January 6, 2021, or the
2016 election -- a platform can freely modify its definition of
``engagement.'' At \emph{t\textsuperscript{2}}, one ``Comment'' is now worth five ``Likes.'' Note
also that social media companies are in the best -- and perhaps only --
position to equalize the amplification of political speech. Their
platforms have been designed by software engineers who hold the key to
recommenders that optimize for engagement. Without a doubt, other logics
of content distribution could be coded into existence.

In other words, social media companies directly contribute to
inequalities of reach through their algorithmic choices. As we have seen
in the preceding section, they also offer features that allow users to
pay for amplification. In doing so, they help reproduce inequalities of
reach that already plague the analogue public sphere. What is more,
platforms have the capacity to ameliorate the pathologies to which they
contribute. By modifying recommendation algorithms, they could ensure a
fairer distribution of speech. For these reasons, it seems reasonable to
pressure social media companies to create a digital public sphere that
better aligns with our democratic ideals, including political equality.

This leaves us with the practical task of determining which measures
would best enable platforms to equalize the reach of online speech. In
what follows, I discuss a practical measure related to recommender
architecture. Certainly, such a measure does not represent the only
possible remedy to the problem of unequal political influence on social
media platforms. Some egalitarians might prefer different recommender
reforms. Furthermore, structural reforms may also be needed to mitigate
inequalities of reach in the digital public sphere. For instance,
antitrust laws could be used to limit platform size and help avoid a
situation where the attention of billions of people is concentrated on
just a few platforms. For instance, Balkin (2021) suggests that
antitrust law could be used to create a social media federalism composed
of many smaller platforms. Alternatively, democratic states could
implement a public grants approach through which they would subsidize
small alternative social media platforms that rank posts using
alternative algorithms. Lastly, states could provide disadvantaged
actors with resources to compete in the digital public sphere, for
instance, by offering free digital literacy education or subsidies that
such actors could use to pay for amplification.

I see no need to reject any of those potential fixes. None of them is in
tension with the measure I propose below. However, until states
implement such structural fixes, social media users will continue to
gravitate toward just a few platforms, where attention is highly
concentrated. In this non-ideal digital public sphere, modifying current
recommendation algorithms offers a technically feasible pathway to
reducing informal inequalities in online political influence, one that
places the financial burden on wealthy platforms that have historically
benefited from injustice.~

Of the algorithmic measures available, recommendation floors represent
the most direct response to unequal opportunity for political influence
by providing each social media user with a meaningful opportunity to
have their voice heard by others, regardless of wealth, digital
literacy, or existing follower base. Concretely, each verified account
would receive a guaranteed minimum number of algorithmically recommended
impressions per set period.\footnote{Specifically, verification would ensure that no single person
  holds several different accounts, and that all accounts belong to
  humans (as opposed to bots).} Although fixing a precise number of
impressions remains hard in an environment where platforms deliberately
withhold recommendation data from the public, we can illustrate how
floors would work through a hypothetical scenario.

Let us imagine a major social media platform where the median number of
impressions for all political content posted by a user during a given
week falls below 500.\footnote{To go beyond this hypothetical scenario and fix a minimum number
  of impressions, data would need to flow from platforms to the
  public. Computer simulations that allow us to observe how social
  dynamics change once recommenders are modified would also be needed.
  Part of my aim, in this paper, is to motivate researchers with
  access to relevant data sets and the technical means to run
  simulations to experiment with recommenders that aim at political
  equality. See, for instance Larooij and Törnberg (2025).} In this context, platforms would ensure that
each account receives, say, 500 impressions for up to 10 political posts
per week, totaling 5000 weekly impressions. In doing so, they would
multiply each user's opportunity to be heard tenfold. The distribution
of these impressions would not be random but interest- and
geography-based. For instance, a post written in Japanese arguing in
favor of creating bike lanes in Tokyo could be distributed to other
Tokyoites who have previously been categorized by machine-learning
models as interested in ``urban planning,'' ``Tokyo,'' ``local politics,''
``biking'' or ``traffic congestion.''

Granted, the exact number of impressions, the number of posts that
qualify for the minimum recommendation, and the appropriate time period
for each platform will vary. Still, recommendation floors would directly
translate a democratic ideal -- equal opportunity for political
influence -- into the currency of algorithmic amplification. Regardless
of how often users post, they would have a more meaningful opportunity
to be heard in the digital public sphere. Recommendation floors would
also encourage substantive participation among users who want to
participate in online political discussions but lack the time or
knowledge to cultivate many followers. A system where individual users
-- rather than individual posts -- are guaranteed a minimum number of
impressions would also prevent gaming by discouraging users from
repeatedly posting low-quality content. Beyond a certain number of
posts, additional posts would receive no special distribution. This
would incentivize users to prioritize quality over quantity.

Importantly, recommendation floors are feasible; they would not burden
social media platforms with developing new technical capacities but
would rather require recalibrating existing ones. On such platforms,
impressions are already tracked, content is already categorized by
topic, and distribution is already interest-based. Of course,
implementing recommendation floors for \emph{qualifying political content}
presupposes that major social media platforms can distinguish between
political and non-political speech. While drawing this distinction is
challenging, it aligns with current moderation policies and practices.
In February 2024, for instance, Meta implemented a new moderation policy
under which it ``won't proactively recommend content about politics on
recommendation surfaces across Instagram and Threads.'' (Instagram 2024)
Here, content about politics is understood as content that mentions
governments, elections, and social topics. Consequently, platforms could
rely on their existing categorization of political speech. To go
further, they could also rely on user self-declaration (i.e., users
flagging their content as political) and community flagging (which is
already employed in content moderation operations).

Admittedly, there are strong moral reasons not to amplify \emph{all}
political speech. As is well known, social media platforms moderate a
significant share of user-generated content, including spam, sexual
content, misinformation, hate speech, and incitement. (Howard 2021) My
suggestion, here, is not that floors should apply to every single piece
of political content. When certain forms of political expression
seriously threaten the rights of listeners, speech can legitimately be
de-amplified or removed from platforms altogether. Concretely, platforms
could maintain their current content moderation practices by removing
harmful content (e.g., violent posts, spam) before it is algorithmically
recommended to users. In other words, only political speech that does
not violate moderation policies should be subject to recommendation
floors. Most pieces of user-generated political content undoubtedly fall
within this category. In the first half of 2022, for instance, Snapchat
estimates that its Violative View Rate was 0.04\%, which means that ``out
of every 10,000 Snap and Story views on Snapchat,'' only ``4 contained
content that violated {[}\ldots{]} policies.'' (Snapchat Transparency Report,
2022)

To sum up, recommendation floors would not achieve perfect equality of
political influence, but they would represent a significant improvement
over the \emph{status quo}. They align with the longstanding democratic ideal
of equal speech, according to which the parity of speakers must be
``recognized and reaffirmed through their equal claim to a public hearing
and others' consideration of what they say.'' (Bejan 2020, 155) By
preventing speakers from falling below the threshold of meaningful
participation, they could also be endorsed by philosophers who, like
Dworkin, fall short of accepting Kolodnian EOPI but nevertheless hold
that democracy requires all citizens to be provided with a ``real chance
to make their voices heard.'' (Dworkin 2011, vii-viii)

While the previous argument focuses on speaker equality, recommendation
floors can also be defended with an appeal to the epistemic interests of
audiences. As Miller notes, the ``low-level amplification provided by
some platforms can be critical for {[}\ldots{]} the mobility of ideas in
democratic discourse.'' (2022, 66) In other words, providing people with
an equal opportunity to be heard not only promotes their interests as
speakers but also helps create a set of competing political ideas from
which they can choose as listeners. That listeners have a strong
interest in being exposed to a diverse set of ideas is a thought deeply
anchored in the Millian free speech tradition. If the mobility of ideas
is hampered, Mill (2008) famously argued, then citizens are less likely
to correct the false beliefs they hold and understand the grounds of
their true opinions, which become salient through discussion and debate.

While floors are an important mechanism for advancing EOPI at the
platform level, other algorithmic reforms could help achieve that goal.
In the remainder of this section, I briefly discuss two such reforms
without attempting to defend them. First, recommendation ceilings would
set a threshold of amplification below which individual instances of
political speech would fall. Concretely, this would translate into a
maximum number of algorithmically recommended impressions for qualifying
political content. Like campaign finance reforms, impression ceilings
would limit the influence of powerful users over the political
decision-making process by hampering their ability to bury the voices of
others.~

Interestingly, Miller (2021) notes that ceilings can also be defended on
the basis of audiences' epistemic interests. Specifically, she argues
that reaching an ever-increasing number of listeners yields decreasing
marginal benefits for speakers. This is at least the case if we center
our reflection on speaker autonomy, conceived as the ability to forge
one's worldview through dialogue and debate. As she explains, what is
\emph{not} needed for autonomy is a very large audience. (2021, 50) To be
autonomous agents, speakers must simply be exposed to divergent
perspectives that allow them to gain critical distance from the ideas
they endorse. Once they are, ``The advantages that additional audience
members provide for freedom of thought begin to decline.'' Here, Miller's
point is that limiting mass amplification serves democratic values such
as equal political influence without seriously impeding speakers'
ability to engage with a wide range of dissenting perspectives.
Arguably, impression ceilings would also promote epistemic competition
-- the mechanism through which socially dominant ideas are challenged by
dissenting views -- which has long been treated as a core mechanism of
self-government by free speech scholars and U.S. Supreme Court Justices
alike.

That said, ceilings would also have negative consequences. In some
instances, elites can use their massive reach to mobilize attention
around neglected issues and discussions. Think, for instance, of a video
containing evidence of police brutality going viral. In this case,
placing a numerical limit on impressions would be bad for justice and
democracy. Furthermore, social media platforms arguably have a duty to
ensure that their users see the speech of elected political leaders who
hold power over them so that they can hold them accountable. If ceilings
prevent them from doing so, then this counts against their
implementation. A related but distinct worry is that voices are
sometimes amplified for legitimate epistemic reasons, as they carry
strong arguments and resonate with many people. If this is the case,
then ceilings could undermine the wisdom of crowds.

However, while the epistemic consequences of algorithmic reforms are
hard to predict, there are reasons to worry that influencers who
dominate the attention economy have incentives to engage in
non-epistemically ideal speech. This is because engagement-based
algorithms often amplify divisive content, and influencers therefore
have an incentive to indulge in this kind of speech. (Milli et al.~2024)
This claim is consistent with the findings of Robertson et al.~(2024),
who suggest that ``online discussions are dominated by a surprisingly
small, extremely vocal, and non-representative minority'' of users who
``stir discontent, spread misinformation and spark outrage online'' (p.
1). There is, therefore, some hope that de-amplifying the currently most
amplified would have good epistemic effects.

Beyond floors and ceilings, one can reasonably worry that, like social
media users who lack money, leisure time, or digital literacy, members
of marginalized groups are currently denied an equal opportunity to be
heard because of stereotypes and structural injustices. For instance, a
recent study of gender differences in Twitter use and influence finds
that women researchers in academic medicine are equally active as men
but have fewer average followers, ``likes,'' and ``retweets'' (Zhu et al.
2019). As its authors explain, this leads to a disparity in visibility
and advancement opportunities, which is especially pernicious in
disciplines like health policy, where ``engagement with health care
decision-makers impacts academic influence, recognition, and promotion''
(1726). The authors also note that their findings put pressure on the
idea that the openness of social media platforms \emph{ipso facto} transforms
them into an equalizing force:

\begin{quote}
Some have hoped that social media would help level the playing field
in academic medicine by giving women an accessible and equitable
platform on which to present themselves. However, our findings---that
women's voices on Twitter appeared to be less influential and have
less reach than men's---suggest that these forums may do little to
improve gender parity and may instead reinforce disparities (1728).
\end{quote}

Granted, this empirical data does not focus on political speech, but the
general point is that recommendation algorithms could be designed to
control for documented biases. If, for instance, women's political
speech circulates much less than men\textquotesingle s on average, then being a woman
could become a weighted factor in the algorithmic recommendation of
political content.

However, one worry about what I propose to call ``affirmative
amplification'' is that it would necessarily rely on indicators of a
user's identity (e.g., self-identification), and that such indicators
might be gamed (e.g., lying about identity to gain more reach). To
prevent gaming, platforms would have to collect information about their
users in ways that would likely violate privacy norms. Some users have
good reasons not to disclose their gender, race, socioeconomic status,
or sexual orientation on social media platforms, and we don't want
platforms to ask. Lastly, in certain jurisdictions, granting
differential treatment to some users based on legally protected
categories might be unlawful.

For these reasons, recommendation floors seem the most promising
algorithmic pathway to EOPI. Ultimately, the key point is that options
exist. Although reimagining the digital public sphere offers a partial
solution to the broader problem of unequal influence in democratic
societies, equalizing online speech might be easier than, say,
restructuring campaign finance or remaking the traditional media
ecosystem. To some extent, online spaces are more malleable than their
analogue counterparts, and algorithmic design offers an unprecedented
opportunity to shape digital fora along egalitarian ideals.

\section*{4. Objections}\label{objections}

One important objection to the idea that social media platforms should
seek to equalize online political influence comes from a libertarian
perspective. According to this view, such platforms are private spaces
owned by entrepreneurs with a stringent property right to exclude users
at will. If so, platforms arguably have the right to deamplify users'
political speech. For instance, Cohen and Cohen (2022, 17) argue that
``nonstate agents ought to enjoy immensely (but not infinitely) stringent
rights to exclude others who seek access to their property, including
for the purposes of expression or communication.'' In their view, social
media platforms resemble malls whose owners can legitimately choose to
``provide customers with a certain atmosphere such as one that excludes
certain political solicitations or only provides certain political
solicitations'' (20). The resulting challenge for egalitarians is to
``show that owners' rights to their spaces must yield to the demands of
others to access such spaces against owners' wishes.''

In response, I suggest that the stringency of owners' proprietary rights
to exclude others depends on the functions the spaces they own serve in
a democratic society. If a private space fulfills an essential public
function, its owners' right to exclude speakers is limited by citizens'
legitimate democratic interests. For example, racists can legally
exclude non-whites from their living rooms, but café owners cannot
control access to their establishments based on race. This is because
people have a moral and legal right against discrimination, one that
protects their democratic interest in equal access to spaces of
gathering and association. Arguably, however, living rooms are not
primarily spaces of gathering and association but of rest and intimacy.
Consequently, their owners must retain the ability to exclude others at
will if they are to remain as such. Normatively speaking, the function
that spaces fulfill is at least equally relevant as the nature of their
ownership. If that were not the case, there would be few reasons to
grant individuals a right to access privately owned open spaces such as
cafés and restaurants. Mainstream social media platforms also fit that
category, as they now fulfill an essential democratic function by
allowing users to discuss and debate a wide range of topics. For better
or worse, such spaces represent digital town squares insofar as they
have captured a large share of the democratic public's collective
attention. If this is so, their owners' interest in excluding users
should be limited by citizens' democratic interests in expressing
themselves and being heard by others. (Theil 2022; Kramer 2021) \emph{Pace}
Cohen and Cohen, private ownership does not immediately nullify these
interests.

Cohen and Cohen's response to this line of reasoning amounts to the
claim that ``the town square argument fails when there are alternative
outlets for expression that can substitute for one's preferred venue to
communicate.'' (2022, 18) While this is true, the strength of this
response depends on how easily speakers can access spaces that attract a
comparable share of the public's collective attention. If you wish to
speak on the busiest square in our town, but I only allow you to do so
in a desolate park on the outskirts, then I am not offering you a
comparable outlet for expression. Moreover, the claim that all social
media users can easily access alternative outlets should be questioned.
As Miller (2021, 66) underlines:

\begin{quote}
For many citizens lacking wealth and resources, speaking at certain
public places or times, or in certain manners, are roughly their only
gateways into broader public discourse and the formation of public
opinion. Letters to the editor of newspapers are often not accepted;
internet blogs and webpages will not be visited unless they are picked
up by amplifying algorithms.
\end{quote}

When speakers lose their preferred tribune, not all of them can easily
find a substitute. This is especially so for members of marginalized
groups who have created spaces of solidarity and resistance by gradually
taking ownership of mainstream social media platforms.\footnote{Black Twitter is an example that comes to mind. See Klassen et al.
  (2021) for a discussion.} In fact, the
substitutes that Cohen and Cohen invite us to consider -- op-eds,
broadcasts, podcasts, or blogs -- are either not as relevant or much
less accessible to the average speaker than social media.

A second objection relates to the impact of equalizing amplification on
social media platforms on listeners' agency. If recommendation floors
deprive users of the ability to curate content, the objection goes, then
this measure arguably runs contrary to users' right to pay attention to
speech of their choosing. In other words, equalizing amplification would
amount to coercing users into being exposed to content they find
uninteresting, offensive, or generally objectionable. In response, I
suggest that recommendation floors are not a serious threat to users'
agency. First, once political posts have met their guaranteed number of
impressions, they would remain eligible for organic network propagation
(e.g., retweets) and engagement-based algorithmic distribution. Second,
users would remain free to follow and block other accounts, repost
content, and allocate more time to specific posts. Certainly, floors
could make it so that users encounter more content with which they do
not wish to engage, but such users would maintain the freedom to scroll
past it. This will surely be judged inconvenient by some. Yet, the costs
of amplifying potentially uninteresting content are modest compared to
those of marginalizing democratic speakers, which is a serious
democratic harm. Third, there are reasons to doubt that current
recommendation practices involve strong user agency and track users'
preferences. To see this, consider once more the distinction between
network propagation and algorithmic recommendation (Narayanan 2023).
With network propagation, information circulates ``as long as users who
see it choose to further propagate it.'' I post, you reshare my post, and
your followers see my post as a result. With algorithmic recommendation,
users see content from users they have not actively chosen to follow
when recommenders predict they will find it engaging. Compared to
network propagation, there is relatively little agency involved in
algorithmic recommendation. With the former, users actively choose to
follow other users; with the latter, content is recommended to them
because of questionable assumptions (e.g., ``User has spent x number of
seconds watching a piece of content A; they are therefore likely to
appreciate a piece of content B'').\footnote{For a critique of the behavioral signals on which current
  recommendation algorithms rely, see Brown (2025).} Even when the recommended content
is engaging, users are not asked whether they primarily want to be
recommended such content or would prefer an alternative recommendation
model. (Schuster and Lazar 2024)

A third objection amounts to the claim that recommendation floors -- or,
for that matter, any measure that redistributes attention -- might hurt
the economic prospects of influencers, especially those who currently
command the most attention on social media platforms. While this might
be the case, my response is that democratic values can justify some
constraints on commercialized political influence. I also worry that
this objection proves too much. After all, regulations that coerce
platforms into limiting the spread of medical misinformation also risk
having an adverse economic effect on their biggest spreaders, but this
consideration does not override the public's interest in being protected
from diseases and pandemics. Furthermore, content creators arguably do
not have the right to demand that social media platforms implement
specific algorithms because this would allow them to generate more
profit.

Another version of this argument directly applies to social media
platforms' economic goals (as opposed to those of content creators). As
platforms must create conditions conducive to monetizing through
advertisements, pressuring them to modify their recommender architecture
might hurt their business model. Without downplaying the importance for
platforms to turn a profit within market-based economies, constraining
market activity is legitimate when doing so promotes important
democratic goals. Moreover, this argument too might prove too much:
health labels on cigarette packs and emission standards for gasoline
cars are also likely to harm the business models of the tobacco and
automotive industries. Yet, we should not reject them for this reason.

\section*{5. Conclusion}\label{conclusion}

In a recent discussion of shadow banning, Tarleton Gillespie (2022)
contends that algorithmic amplification and reduction should be at the
forefront of scholarly reflections on platform accountability. He
explains, ``We do not know whether we can trust platforms to engage in
these reduction practices thoughtfully, in ways that produce a robust
but fairer public sphere.'' (9) Certainly, platforms are designed to make
it seem \emph{as if} content encountered by users spontaneously emerges from
other users' minds. In reality, each platform implements what Gillespie
dubs a ``logic of circulation.'' Every day, policy managers and software
engineers make decisions about what gets seen and what does not. As
discussed, the logic of circulation that currently predominates on
mainstream social media platforms is that of engagement optimization;
users are presented with content that arouses strong emotions and
reactions, which leads them to spend more time on platforms. These
reactions generate data exhausts that translate into advertising
revenues.

My main objective in this article has been to help define a
countervailing logic of circulation. Instead of optimizing for
engagement, we should optimize for democratic values, including
political equality. Specifically, I have argued that political speakers
should be provided with an equal opportunity to be heard and proposed an
algorithmic reform -- recommendation floors -- that would move the
digital public sphere closer to that ideal. Without a doubt, democratic
equality is not the sole value to consider when designing recommender
systems. Ultimately, I do not pretend to have identified the single best
way to balance all relevant democratic values in the digital public
sphere. By discussing the implications of contemporary egalitarianism
for platform architecture, I more modestly hope to incite readers to
raise for themselves the central question that animates my reflection:
of all possible logics of circulation that could govern communicative
practices in the digital public sphere, which one should prevail? By
raising this question, the democratic public can begin to reappropriate
digital communication spaces and combat the self-serving logic of
engagement optimization.

\section*{References}
\begin{hangingbib}
\item Archer, Alfred, Amanda Cawston, Benjamin Matheson, and Machteld Geuskens. 2020. ``Celebrity, Democracy, and Epistemic Power.'' \emph{Perspectives on Politics} 18 (1): 27--42. \url{https://doi.org/10.1017/S1537592719002615}.

\item Balkin, Jack M. 2021. ``How to Regulate (and Not Regulate) Social Media.'' \emph{Journal of Free Speech Law}, no. 71. \url{https://doi.org/10.2139/ssrn.3484114}.

\item Bejan, Teresa M. 2020. ``Free Expression or Equal Speech?'' \emph{Social Philosophy and Policy} 37 (2): 153--69. \url{https://doi.org/10.1017/S0265052521000091}.

\item Bonotti, Matteo, and Jonathan Seglow. 2022. ``Freedom of Speech: A Relational Defence.'' \emph{Philosophy \& Social Criticism} 48 (4): 515--29. \url{https://doi.org/10.1177/01914537211073782}.

\item Bousquet, Chris. 2025. ``Limiting the Right to Moderate: Political Equality, Social Media, and Viewpoint-Based Moderation.'' \emph{Ethics and Information Technology} 27 (4): 62. \url{https://doi.org/10.1007/s10676-025-09875-w}.

\item Brown, Etienne. 2025. ``Recommended Selves: Authenticity and Algorithmic Filtering.'' \emph{Journal of the American Philosophical Association}, September 22, 1--20. \url{https://doi.org/10.1017/apa.2025.10009}.

\item Cohen, Andrew, and Andrew Cohen. 2022. ``The Possibility and Defensibility of Nonstate `Censorship.'\,'' In \emph{New Directions in the Ethics and Politics of Speech}, 1st ed., by J.P. Messina. Routledge. \url{https://doi.org/10.4324/9781003240785-2}.

\item Dworkin, Ronald. 2011. ``Foreword.'' In \emph{Extreme Speech and Democracy}. Oxford University Press.

\item Gillespie, Tarleton. 2022. ``Do Not Recommend? Reduction as a Form of Content Moderation.'' \emph{Social Media + Society} 8 (3): 205630512211175. \url{https://doi.org/10.1177/20563051221117552}.

\item Howard, Jeffrey W. 2021. ``Extreme Speech, Democratic Deliberation, and Social Media.'' In \emph{The Oxford Handbook of Digital Ethics}, 1st ed., edited by Carissa Véliz. Oxford University Press. \url{https://doi.org/10.1093/oxfordhb/9780198857815.013.10}.

\item Instagram. 2024. \emph{Continuing Our Approach to Political Content on Instagram and Threads}. February 9.

\item Justia Law. n.d. ``Kovacs v. Cooper, 336 U.S. 77 (1949).'' Accessed June 13, 2023. \url{https://supreme.justia.com/cases/federal/us/336/77/}.

\item Klassen, Shamika, Sara Kingsley, Kalyn McCall, Joy Weinberg, and Casey Fiesler. 2021. ``More than a Modern Day Green Book: Exploring the Online Community of Black Twitter.'' \emph{Proceedings of the ACM on Human-Computer Interaction} 5 (CSCW2): 1--29. \url{https://doi.org/10.1145/3479602}.

\item Kolodny, Niko. 2014. ``Rule Over None II: Social Equality and the Justification of Democracy.'' \emph{Philosophy \& Public Affairs} 42 (4): 287--336. \url{https://doi.org/10.1111/papa.12037}.

\item Kramer, Matthew H. 2021. \emph{Freedom of Expression as Self-Retraint}. First edition. Oxford Scholarship Online. Oxford University Press. \url{https://doi.org/10.1093/oso/9780198868651.001.0001}.

\item Larooij, Maik, and Petter Törnberg. 2025. ``Can We Fix Social Media? Testing Prosocial Interventions Using Generative Social Simulation.'' Version 1. Preprint, arXiv. \url{https://doi.org/10.48550/ARXIV.2508.03385}.

\item Lazar, Seth. 2025. ``Governing the Algorithmic City.'' \emph{Philosophy \& Public Affairs} 53 (2): 102--68. \url{https://doi.org/10.1111/papa.12279}.

\item Mill, John Stuart. 2008. \emph{On Liberty, Utilitarianism and Other Essays}. Oxford: Oxford University Press.

\item Miller, Erin. 2021. ``Amplified Speech.'' \emph{Cardozo Law Review} 43 (1).

\item Milli, Smitha, Micah Carroll, Yike Wang, Sashrika Pandey, Sebastian Zhao, and Anca Dragan. 2024. ``Engagement, User Satisfaction, and the Amplification of Divisive Content on Social Media.'' \emph{Knight First Amendment Institute at Columbia University Essays and Scholarship}, January 3. \url{https://knightcolumbia.org/content/engagement-user-satisfaction-and-the-amplification-of-divisive-content-on-social-media}.

\item Robertson, Claire E., Kareena S. Del Rosario, and Jay J. Van Bavel. 2024. ``Inside the Funhouse Mirror Factory: How Social Media Distorts Perceptions of Norms.'' \emph{Current Opinion in Psychology} 60 (December): 101918. \url{https://doi.org/10.1016/j.copsyc.2024.101918}.

\item Schuster, Nick, and Seth Lazar. 2024. ``Attention, Moral Skill, and Algorithmic Recommendation.'' \emph{Philosophical Studies}, ahead of print, January 22. \url{https://doi.org/10.1007/s11098-023-02083-6}.

\item Shiffrin, Seana V., \emph{Speech Matters: On Lying, Morality, and the Law}. 2014. Carl G. Hempel Lecture Series. Princeton: Princeton University Press.

\item Smith, Leonie, and Alfred Archer. 2020. ``Epistemic Injustice and the Attention Economy.'' \emph{Ethical Theory and Moral Practice} 23 (5): 777--95. \url{https://doi.org/10.1007/s10677-020-10123-x}.

\item ``Snapchat Transparency Report \textbar{} Snapchat Transparency.'' 2022. \url{https://values.snap.com/privacy/transparency}.

\item Stray, Jonathan, Alon Halevy, Parisa Assar, et al.~2022. \emph{Building Human Values into Recommender Systems: An Interdisciplinary Synthesis}. \url{https://doi.org/10.48550/ARXIV.2207.10192}.

\item Theil, Stefan. 2022. ``Private Censorship and Structural Dominance: Why Social Media Platforms Should Have Obligations To Their Users Under Freedom of Expression.'' \emph{The Cambridge Law Journal} 81 (3): 645--72. \url{https://doi.org/10.1017/S0008197322000484}.

\item Thorburn, Luke. 2022. ``How Platform Recommenders Work.'' \emph{Understanding Recommenders}, November 23. \url{https://medium.com/understanding-recommenders/how-platform-recommenders-work-15e260d9a15a}.

\item Zhu, Jane M., Arthur P. Pelullo, Sayed Hassan, Lillian Siderowf, Raina M. Merchant, and Rachel M. Werner. 2019. ``Gender Differences in Twitter Use and Influence Among Health Policy and Health Services Researchers.'' \emph{JAMA Internal Medicine} 179 (12): 1726. \url{https://doi.org/10.1001/jamainternmed.2019.4027}.
\end{hangingbib}

\end{document}